\documentclass[%
aip,
jcp,%
amsmath,amssymb,
reprint,%
]{revtex4-1}

\usepackage{graphicx}% Include figure files
\usepackage{dcolumn}% Align table columns on decimal point
\usepackage{bm}% bold math
\usepackage{xcolor}
\begin{document}

\preprint{AIP/123-QED}

\title[Viscoelastic Phase Separation Model for Ternary Polymer Solutions]{Viscoelastic Phase Separation Model for Ternary Polymer Solutions}% Force line breaks with \\
%\thanks{Footnote to title of article.}

\author{Kenji Yoshimoto}
%\email{yoshimoto1@cheme.kyoto-u.ac.jp}

\author{Takashi Taniguchi}%
\email{taniguchi@cheme.kyoto-u.ac.jp}
\affiliation{Department of Chemical Engineering, Kyoto University, Kyoto 615-8510, Japan}%
%\affiliation{Department of Chemical Engineering, Graduate School of Engineering, Kyoto University, Kyoto 615-8510, Japan}%
% \homepage{http://www.Second.institution.edu/~Charlie.Author.}
%\affiliation{%
%Second institution and/or address%\\This line break forced% with \\
%}%

\date{\today}% It is always \today, today,

\begin{abstract}
When a polymer solution undergoes viscoelastic phase separation, the polymer-rich phase forms a network-like structure even if it is a minor phase. This unique feature is induced by polymer dynamics, which are constrained by the temporal entanglement of polymer chains. The fundamental mechanisms of viscoelastic phase separation have already been elucidated by theory and experiments over the past few decades; however, it is not yet well understood how viscoelastic phase separation occurs in multicomponent polymer solutions. Here, we construct a new viscoelastic phase separation model for ternary polymer solutions that consist of a polymer, solvent, and nonsolvent. Our simulation results reveal that a network-like structure is formed in the ternary bulk system through a phase separation mechanism similar to that observed in binary polymer solutions. A difference in dynamics is also found in that the solvent, whose affinity to the polymer is similar to that to the nonsolvent, moves freely between the polymer-rich and water-rich phases during phase separation. These findings are considered important for understanding the phase separation mechanism of ternary mixtures often used in the manufacturing of polymeric separation membranes.
\end{abstract}

\pacs{82.60. Lf, 83.80. Rs, 83.60. Bc}% 82.60.Lf: phase separation, 83.80.Rs: polymer solution, 83.60.Bc: linear viscoelasticity,
\keywords{Polymer, Ternary Solution, Phase Separation, Viscoelastic, Model, Simulations}%Use showkeys class option if keyword
\maketitle

%\begin{quotation}
%The ``lead paragraph'' is encapsulated with the \LaTeX\
%\verb+quotation+ environment and is formatted as a single paragraph before the first section heading.
%(The \verb+quotation+ environment reverts to its usual meaning after the first sectioning command.)
%Note that numbered references are allowed in the lead paragraph.
%
%The lead paragraph will only be found in an article being prepared for the journal \textit{Chaos}.
%\end{quotation}

%\section{\label{sec:level1}First-level heading:\protect\\ The line
%break was forced \lowercase{via} \textbackslash\textbackslash}

\section{\label{sec:intro}Introduction}
When a binary liquid mixture undergoes phase separation, the minor component usually forms droplets in the matrix of the major component. However, if the mixture is composed of solvent and polymer and if the polymer chains are long enough to be entangled, the polymer solution may exhibit unique phase separation, termed viscoelastic phase separation.\cite{doi86,tanaka00,taniguchi04} For example, in the early stage of viscoelastic phase separation, the major component, {\it i.e. }, the solvent, forms small droplets.\cite{tanaka96,onuki97} In the late stage, the polymer-rich phase forms a network-like continuous structure even if it is the minor phase.\cite{taniguchi96}
These characteristics are due to a large difference in dynamic behavior between the solvent and the polymer.\cite{tanaka96} In the case of a binary liquid, each component diffuses down the concentration gradient, with speeds nearly the same between the two components.
%Editor: Please ensure that the intended meaning has been maintained in this edit.
In viscoelastic phase separation, however, the entangled polymer chains move much more slowly than the solvent. In addition, the dynamic mechanical response differs between the polymer and the solvent. The entangled polymer deforms instantaneously as an elastic body does but also flows like a liquid over a period much longer than the relaxation time.
%Editor: Please ensure that the intended meaning has been maintained in this edit.
These differences, which are referred to as dynamic asymmetries, may affect the mechanism and morphology of phase separation through tight coupling with concentration diffusion.\cite{tanaka00,tanaka96}

In addition to the binary polymer solution, phase separation of the {\it ternary} polymer solution, which consists of polymer, solvent, and nonsolvent ({\it e.g.}, water), is particularly important when preparing a polymer membrane. For example, in the nonsolvent-induced phase separation (NIPS) process, which is widely used for the manufacture of polymer membranes, the polymer is dissolved in a solvent that is soluble both in the polymer and in water. By immersing the polymer solution in water, the solvent is partially replaced by the water. The resulting {\it ternary} polymer solution is thermodynamically unstable due to a strong repulsion between the water and the polymer and immediately separates into water-rich (= major) and polymer-rich (= minor) phases. The water-rich phase forms small droplets that become pores, while the polymer-rich phase forms a network-like structure as a matrix. Both phases are essentially similar to those observed in the viscoelastic phase separation of a binary polymer solution.

To provide insight into the phase separation mechanism in the NIPS process, some phase separation models have been proposed for the {\it ternary} polymer solution.\cite{zhou06,mino15,tree17} Note, however, that the viscoelastic nature of the entangled polymer is not explicitly considered in any models. For example, the free energy of the {\it ternary} polymer solution has been described as a sum of the Flory-Huggins mixing energy and the interfacial energy, neglecting the elastic energy generated by deformation of the entangled polymer. Additionally, the relaxation of the entangled polymer is not considered in the existing models. Indeed, it has been demonstrated with binary polymer solutions that both elastic and relaxation behavior of the entangled polymer plays a crucial role in the formation of the frozen polymer-rich phase at the early stage and of the network-like polymer-rich structure at the late stage.\cite{taniguchi96,tanaka97,araki01,taniguchi04}

In this paper, we develop a basic model for the phase separation of a {\it ternary} polymer solution that explicitly includes the elastic and relaxation characteristics of the entangled polymer. Our approach is based on the theory of viscoelastic phase separation developed for binary polymer solutions.\cite{doi86,taniguchi96,tanaka97,araki01}
First, we describe the free energy of the {\it ternary} polymer solution as a functional of the volume fractions of each component and the deformation tensor of the entangled polymer chains (Sec. \ref{sec2:variable} and \ref{sec2:thermo}). Then, we prepare a set of equations for evolving the volume fractions and the deformation tensor over time (Sec. \ref{sec2:evolution}). The coupling between the volume fractions and the deformation tensor occurs through the polymer velocity, which is determined from a balance among the thermodynamic, hydrodynamic, and elastic forces. Three different expressions for the polymer velocity are examined: a thermodynamic term only (diffusion case), thermodynamic + hydrodynamic terms (viscous case), and thermodynamic + hydrodynamic + elastic terms (viscoelastic case) (Sec. \ref{sec:sim}). The first two cases are representatives of the previous models, and the last case is the focus of development in this study. Numerical calculations are performed to simulate the spinodal decomposition of a bulk {\it ternary} polymer solution and the change in morphology of a square-shaped polymer-rich phase immersed into a water-rich bath. By comparing the results obtained from the three cases, we highlight some important aspects of the viscoelasticity of an entangled polymer on the phase separation of a {\it ternary} polymer solution.

\subsection{\label{sec2:variable}Field variables: $\phi$ and ${\bm W}$}
A ternary mixture of polymer, solvent, and water is described by three volume fractions: $\phi_{\rm p}$, $\phi_{\rm s}$, and $\phi_{\rm w}$. The subscripts ${\rm p}$, ${\rm s}$, and ${\rm w}$ denote polymer, solvent, and water, respectively. Each volume fraction varies with position ${\bm r}$ and time $t$, whereas the sum of the three volume fractions is always unity due to incompressibility;
\begin{eqnarray}
  \sum_{\alpha={\rm p},{\rm s},{\rm w}}\phi_{\alpha} = 1.
  \label{eq:incomp}
\end{eqnarray}
In this paper, subscript $\alpha$ (or $\beta$) is used to denote one of the three components: polymer, solvent, and water.

\begin{figure}
\includegraphics[scale=0.28]{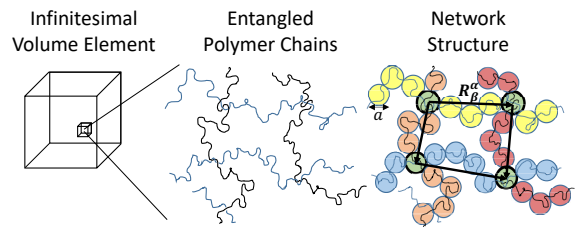}%
 \caption{\label{fig1} Schematic representation
 of the temporal network structure of entangled polymer chains.}
\end{figure}

Polymer chains are assumed to be long enough to have temporal entanglements, as schematically illustrated in Fig. \ref{fig1}. The entangled polymer chains form a temporal network structure that behaves like an elastic object for instantaneous, infinitesimal deformation. In contrast, the network structure flows like liquid under a constant load as a result of the stretch and orientation relaxations of entangled polymer chains by reptation motion and a constraint release mechanism.\cite{doi86} Such viscoelastic behavior of polymer entanglements can be included in the model by introducing another field variable, the so-called conformation tensor ${\bm W}$.\cite{taniguchi96}
A microscopic definition of ${\bm W}$ is given as follows. Let us consider an infinitesimal volume element that contains $n_{\rm p}$ $(\gg 1)$ polymer chains. As schematically shown on the right-hand side of Fig.~\ref{fig1}, all polymer chains are assumed to be Gaussian composed of $N$ statistical segments with a size of $a$. The $\ell$-th polymer chain ($\ell=1, \cdots, n_{\rm p}$) has $Z_{\ell}$ $(\gg 1)$ entangled segments whose positions are denoted by ${\bm r}_1^{\ell}, ..., {\bm r}_{Z_{\ell}}^{\ell}$. The vectors connecting two successive entangled segments, ${\bm R}_{m}^{\ell}$ ($ = {\bm r}_{m+1}^{\ell} - {\bm r}_{m}^{\ell}$, where $m = 1, ..., Z_{\ell}-1)$, are considered as a frame of the network structure. At equilibrium, the average length of ${\bm R}_{m}^{\ell}$ is $a\sqrt{N_{\rm e}}$, where $N_{\rm e}$ is the average number of segments existing between the two adjacent entangled segments along a chain. The normalized dyadic tensor ${\bm R}_{m}^{\ell}{\bm R}_{m}^{\ell}/a^2N_{\rm e}$ represents the change in shape of the network structure from equilibrium. The average of ${\bm R}_{m}^{\ell}{\bm R}_{m}^{\ell}/a^2N_{\rm e}$ over all the entangled segments and chains existing in the same infinitesimal volume element is defined as the conformation tensor ${\bm W}$;
\begin{eqnarray}
  {\bm W} \equiv \frac{1}{n_{\rm p}}\left[\sum_{\ell=1}^{n_{\rm p}}\frac{1}{(Z_{\ell}-1)}\sum_{m=1}^{Z_{\ell}-1}\frac{{\bm R}_{m}^{\ell}{\bm R}_{m}^{\ell}}{3a^2{N_{\rm e}}}\right].
  \label{eq:w}
\end{eqnarray}
The right-hand side of Eq. (\ref{eq:w}) becomes an identity tensor ${\bm I}$ at equilibrium.
Eq. (\ref{eq:w}) provides microscopic insight into the origin of polymer viscoelasticity. However, calculation of ${\bm W}$ from Eq. (\ref{eq:w}) is computationally demanding since it requires an explicit description of all polymer chains and entanglements.
Alternatively, one can treat ${\bm W}$ as a continuous field variable and evolve it through a constitutive equation.\cite{larson13} In this approach, microscopic details of the polymer chains are discarded, whereas any local deformation of the network structure can be described by the same macroscopic constitutive relation arising from numerous entanglements of polymer chains. As described below, we take the latter approach, treating ${\bm W}$ as a continuous, frame-invariant deformation tensor and evolving ${\bm W}$ by a simple viscoelastic constitutive equation.\cite{larson13}

\subsection{\label{sec2:thermo}Thermodynamic properties}
Hereafter, we make all variables and equations dimensionless to generalize the model. For a ternary system with constant temperature $T$ and volume $V$, the free energy $F$ can be described as a functional of the volume fractions and conformation tensor;
\begin{eqnarray}
  F[\phi_{\rm p}, \phi_{\rm s}, \phi_{\rm w}, {\bm W}] = \int_Vd{\bm r}\left(
  f_{\rm mix} + f_{\rm int} + f_{\rm ela}\right),
  \label{eq:F}
\end{eqnarray}
where $\phi_{\rm w} = 1 - \phi_{\rm p} - \phi_{\rm s}$.
The first integrand in the right-hand side of Eq. (\ref{eq:F}), $f_{\rm mix}$, represents a mixing free energy as\cite{flory42}
\begin{eqnarray}
  f_{\rm mix} =&& \frac{\phi_{\rm p}}{N}\text{ln}\phi_{\rm p}+\phi_{\rm s}\text{ln}\phi_{\rm s}+\phi_{\rm w}\text{ln}\phi_{\rm w} \nonumber\\
  &&+\chi_{\rm ps}\phi_{\rm p}\phi_{\rm s} + \chi_{\rm sw}\phi_{\rm s}\phi_{\rm w} + \chi_{\rm wp}\phi_{\rm w}\phi_{\rm p},
  \label{eq:fmix}
\end{eqnarray}
where $\chi_{\alpha\beta}$ is the Flory-Huggins parameter for the affinity between two different components, $\alpha$ and $\beta$ ($\ne \alpha$).
The second integrand $f_{\rm int}$ is the interfacial energy given by
\begin{eqnarray}
  f_{\rm int} = &&\frac{\kappa_{\rm ps}}{2}\vert {\bm \nabla}\left(\phi_{\rm p} - \phi_{\rm s} \right) \vert ^2 + \frac{\kappa_{\rm sw}}{2}\vert {\bm \nabla}\left(\phi_{\rm s} - \phi_{\rm w} \right) \vert ^2 \nonumber\\
  &&+\frac{\kappa_{\rm wp}}{2}\vert {\bm \nabla}\left(\phi_{\rm w} - \phi_{\rm p} \right) \vert ^2, \label{eq:fint}
\end{eqnarray}
where $\kappa_{\alpha\beta}$ is an interfacial parameter for a pair of components $\alpha$ and $\beta$ ($\ne \alpha$).
The last integrand $f_{\rm ela}$ represents elastic energy arising from deformation of the temporal network composed of entangled polymer chains. Among several different formulae proposed for $f_{\rm ela}$, we select the one that contains only the shear modulus of the polymer solution;\cite{taniguchi96}
\begin{eqnarray}
  f_{\rm ela} = \frac{G}{4}\left({\bm W}-{\bm I}\right):\left({\bm W}-{\bm I}\right),
  \label{eq:fela}
\end{eqnarray}
where ${\bm I}$ is the identity tensor.
The shear modulus $G$ is assumed to vary with the polymer concentration based on the scaling theory;\cite{onuki90}
\begin{eqnarray}
  G = G_{\rm 0}\phi_{\rm p}^3.
  \label{eq:g}
\end{eqnarray}
%\textcolor{red}{
The scaling of Eq. (\ref{eq:g}) stems from the characteristic length
between the adjacent entangled segments, $\xi$. According to the book of
de Gennes\cite{degennes79}, the shear modulus of entangled polymer
solution can be expressed as $G\sim k_{\rm B}T/\xi^3$. Since $\xi$
scales with $1/\phi_{\rm p}$, $G$ is inversely proportional to
$\phi_{\rm p}^3$ as shown in Eq. (\ref{eq:g}).
%}
The coefficient $G_0$ will be used as a parameter to characterize the elasticity of the polymer solution.

The chemical potential ${\mu_{\alpha}}$ ($\alpha=$ p, s) and the stress tensor ${\bm \sigma}$ are derived from the functional derivative of the free energy $F$ in Eq.~(\ref{eq:F});\cite{milner93}
\begin{eqnarray}
  \mu_{\alpha} &=& \frac{\delta F}{\delta \phi_{\alpha}} = \frac{\partial f}{\partial \phi_{\alpha}} - {\bm \nabla}\cdot\frac{\partial f}{\partial \left({\bm \nabla}\phi_{\alpha}\right)},
  \label{eq:mu}\\
  {\bm \sigma} &=& 2{\bm W} \cdot \frac{\delta F}{\delta {\bm W}} = 2G\bm{W} \cdot \left({\bm W} - {\bm I}\right),
  \label{eq:sig_e}
\end{eqnarray}
where $f \equiv f_{\rm mix} + f_{\rm int} + f_{\rm ela}$. The detailed expressions of $\mu_{\alpha}$ are summarized in Appendix \ref{app:mu}.

\subsection{\label{sec2:evolution}Time evolution of the field variables}
The volume fractions are evolved with the equations of continuity;
\begin{eqnarray}
  \frac{\partial \phi_{\rm p}}{\partial t} = - {\bm \nabla}\cdot\left(\phi_{\rm p}{\bm v}_{\rm p}\right), \quad  \frac{\partial \phi_{\rm s}}{\partial t} = - {\bm \nabla}\cdot\left(\phi_{\rm s}{\bm v}_{\rm s}\right). 
\label{eq:con3}
\end{eqnarray}
Note that $\phi_{\rm w}$ is determined by the compressibility condition, Eq. (\ref{eq:incomp}).
%\textcolor{red}{
In this study, the random noise term arising from the thermal
fluctuations\cite{fukawatase15} is not included in Eq. (\ref{eq:con3})
so as to clarify the viscoelastic effects on the domain dynamics at the
late stage of phase separation. In general, the thermal fluctuations
have little influence on the domain dynamics after the formation of
sharp interfaces\cite{Brown93}.
%}
The velocities of the polymer and solvent, ${\bm v}_{\rm p}$ and ${\bm v}_{\rm s}$, respectively, can be obtained by solving the equations of motion derived based on the concept of {\it stress division}\cite{onuki94}. The resulting expressions are written as
\begin{eqnarray}
  {\bm v}_{\rm p} &=& {\bm v} - \frac{1}{\phi_{\rm p}}\left\{L_{\rm pp}\left({\bm \nabla}{\mu}_{\rm p} - \frac{{\bm \nabla}\cdot{\bm \sigma}}{\phi_{\rm p}}\right)
  + L_{\rm ps}{\bm \nabla}{\mu}_{\rm s}\right\},\quad \label{eq:vp1}\\
  {\bm v}_{\rm s} &=& {\bm v} - \frac{1}{\phi_{\rm s}}\left\{L_{\rm sp}\left({\bm \nabla}{\mu}_{\rm s} - \frac{{\bm \nabla}\cdot{\bm \sigma}}{\phi_{\rm s}}\right)
  + L_{\rm ss}{\bm \nabla}{\mu}_{\rm s}\right\},\quad \label{eq:vs1}
\end{eqnarray}
where $L_{\alpha\beta}$ is the component of the transport coefficient matrix ${\bm L}$
and where ${\bm v}$ is the mean velocity defined as
\begin{eqnarray}
  {\bm v} = {\bm v}_{\rm p}\phi_{\rm p} + {\bm v}_{\rm s}\phi_{\rm s} + {\bm v}_{\rm w}\phi_{\rm w}.
  \label{eq:v}
\end{eqnarray}
The mean velocity ${\bm v}$ satisfies the Stokes equation;
\begin{eqnarray}
  0=-{\bm \nabla}p - \sum_{i={\rm p,s}}\phi_{i}{\bm \nabla}\mu_{i} + \eta\Delta{\bm v} + {\bm \nabla}\cdot{\bm \sigma},
  \label{eq:stokes}
\end{eqnarray}
where $p$ and $\eta$ denote the pressure and viscosity of the ternary mixture, respectively. In this study, we simplify Eqs.~(\ref{eq:vp1}) and (\ref{eq:vs1}) by assuming that the diagonal components of ${\bm L}$ are constants and the off-diagonal components are zero;
\begin{eqnarray}
  {\bm v}_{\rm p} &=& {\bm v} - \frac{L_{\rm pp}}{\phi_{\rm p}}\left({\bm \nabla}{\mu}_{\rm p} - \frac{{\bm \nabla}\cdot{\bm \sigma}}{\phi_{\rm p}}\right),\quad \label{eq:vp2}\\
  {\bm v}_{\rm s} &=& {\bm v} - \frac{L_{\rm ss}}{\phi_{\rm s}}{\bm \nabla}{\mu}_{\rm s}.\quad \label{eq:vs2}
\end{eqnarray}
Equation~(\ref{eq:vp2}) has exactly the same expression as that derived for a polymer-solvent binary system.\cite{taniguchi96} A new addition to the ternary system is Eq.~(\ref{eq:vs2}), which allows the solvent to move separately from the polymer.

To evolve the conformation tensor simultaneously with the volume fractions described above, we employ a constitutive equation, the so-called upper-convected Maxwell equation,\cite{larson13}
\begin{eqnarray}
 \frac{D {\bm W}}{D t} =  \left({\bm \nabla}{\bm v}_{\rm p}\right)\cdot{\bm W} + {\bm W}\cdot\left({\bm \nabla}{\bm v}_{\rm p}\right)^{\rm T} - \frac{1}{\tau}\left({\bm W} - {\bm I}\right),\quad
  \label{eq:Wt}
\end{eqnarray}
where $D/Dt$ represents the substantial derivative (=$\partial/\partial t + {\bm v}_{\rm p}\cdot{\bm \nabla}$) and $\left({\bm \nabla}{\bm v}_{\rm p}\right)^{\rm T}$ denotes the transpose of the velocity gradient tensor.
The relaxation time $\tau$ is assumed to have the same dependency on $\phi_{\rm p}$ as the shear modulus defined in Eq. (\ref{eq:g});
\begin{eqnarray}
  \tau = \tau_{\rm 0}\phi_{\rm p}^3,
  \label{eq:tau}
\end{eqnarray}
with a constant $\tau_{\rm 0}$.

\section{\label{sec:sim}Simulations}
\begin{table}
\caption{\label{tab:para} Simulation parameters.}
\begin{ruledtabular}
\begin{tabular}{lcr}
Parameter & Value\\
\hline
Number of statistical segments per chain: $N$ & 10\\ 
Flory-Huggins parameters: $\chi_{\rm ps}$, $\chi_{\rm sw}$, $\chi_{\rm wp}$  & 0.4, 0.3, 2.3\\
Initial volume fractions: $\phi_{\rm p0}$, $\phi_{\rm s0}$, $\phi_{\rm w0}$  & 0.15, 0.60, 0.25\\
Interfacial parameters: $\kappa_{\rm ps}$, $\kappa_{\rm sw}$, $\kappa_{\rm wp}$& 1.0, 1.0, 1.0\\
Grid size: $d$ & 2.0\\
Box length: $D$ & 512\\
Time step: $\Delta t$ & 0.025\\
%Maximum time: $t_{\rm max}$ & \\
Transport coefficients: $L_{\rm pp}$, $L_{\rm ss}$ & 1.0, 1.0\\
Viscoelastic parameters: $\tau_0$, $G_0$, $\eta$ & 10, 100, 0.1
\end{tabular}
\end{ruledtabular}
\end{table}
%
% phase_diagram3 /Users/admin/Documents/MATLAB/binodal2_N10_20200301
% frame width 1 -> 2, note: /Users/admin/Documents/MATLAB/alchemyst-ternplot-9c72b90/ -> ternaxes.m, ternlabel.m, termplot.m need to be modified
\begin{figure}
  \includegraphics[scale=0.4]{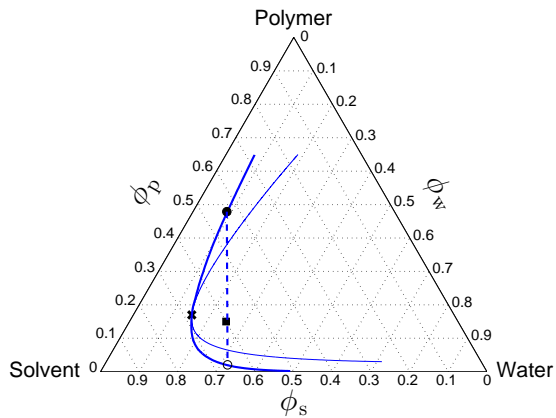}%
\caption{\label{fig:phase_diagram} Phase diagram for the ternary system. The thin and thick lines represent the spinodal and binodal curves, respectively, and they merge on the critical point ($\times$). The simulation starts from a homogeneously mixed state ($\blacksquare$) and eventually reaches equilibrium composed of polymer-rich ($\bullet$) and water-rich ($\circ$) phases. The final compositions of these phases can be connected with a straight line (dashed line) passing through the initial composition.}
\end{figure}
\begin{figure}
  \includegraphics[scale=0.30]{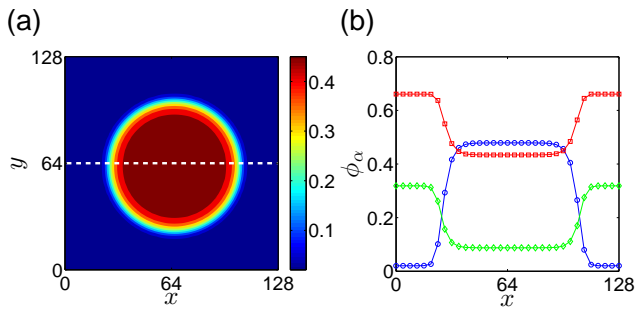}% Here is how to import EPS art
  \caption{\label{fig:equi_comp} (color online) Polymer-rich domain at equilibrium. (a) Two-dimensional plot of the polymer volume fraction $\phi_{\rm p}$. Color is assigned to each square grid based on the value of $\phi_{\rm p}$. For example, red and blue colors represent polymer-rich and water-rich phases, respectively. (b) Profile of the volume fraction along the dashed line in (a). All three volume fractions, $\phi_{\rm p}$, $\phi_{\rm s}$, and $\phi_{\rm w}$, are shown here with the symbols of circle, square, and diamond, respectively. }
\end{figure}
\subsection{\label{sec2:parameter}Parameters}
The parameters used in the simulations are summarized in Table \ref{tab:para}. The Flory-Huggins parameters $\chi_{\alpha\beta}$ are chosen from the literature\cite{matsuyama99} to represent a ternary mixture composed of polyvinylidene difluoride (PVDF), dimethylformamide (DMF), and water. Note that $\chi_{\rm ps}$ and $\chi_{\rm sw}$ depend on the concentration.\cite{matsuyama99} However, for brevity, they are assumed to be constant. The number of statistical segments per chain, $N$, is set at a relatively small number, 10, to increase numerical stability and to perform large-scale simulations.

Figure \ref{fig:phase_diagram} illustrates a phase diagram of the
ternary system with $N$ and $\chi_{\alpha\beta}$ given in Table
\ref{tab:para}. The binodal and spinodal curves are drawn using the
mixing free energy density given in Eq. (\ref{eq:fmix}).
%\textcolor{red}{
The initial polymer volume fraction, $\phi_{\rm p0}$ is set to be 0.15,
which a typical value used in the NIPS process\cite{matsuyama99}. The
initial volume fraction of the water, $\phi_{\rm w0}$, is arbitrary
chosen to be 0.25, assuming that $\sim30\%$ of the solvent in the bulk
polymer solution is instantaneously substituted with the water. With the
given initial composition, the ternary bulk system is already located at
the inside of spinodal region so that it is spontaneously decomposed
into polymer-rich and water-rich phases.
%}
The final compositions of the two phases, $\phi'_{\alpha,{\rm eq}}$ and $\phi''_{\alpha,{\rm eq}}$, are obtained from the fully equilibrated morphology shown in Fig. {\ref{fig:equi_comp}}.

The interfacial parameters $\kappa_{\alpha\beta}$ are set arbitrarily to 1.0. Then, the grid size $d$ is adjusted to 2.0, with which the interfacial regions are expressed by 5 or 6 grid points [see Fig. \ref{fig:equi_comp}(b)]. For the time-related parameters, the transport coefficients, $L_{\rm pp}$ and $L_{\rm ss}$, are set to unity. Then, the time step $\Delta t$ is increased to the maximum, 0.025, with which the iterative calculations can be performed stably over a long course of simulation time.

Finally, the two viscoelastic parameters, $\tau_0$ and $G_0$, are set to 10 and 100, respectively. The viscosity of the ternary system, $\eta$, is set at 0.1, which is estimated from $\eta=G\tau$ with $\phi_{\rm p}=\phi_{\rm p0}$.

\subsection{Three typical cases}
To grasp some fundamental behaviors of phase separation in the ternary system, we perform simulations for the following three cases. The same equations and parameters are used in all cases, except that expressions for the polymer and solvent velocities are modified depending on the cases.\newline
(I) {\it Diffusion} case. The velocity of component $\alpha$ (=p,s) is induced only by the gradient of the chemical potential of the same component;
\begin{eqnarray}
  {\bm v}_{\alpha} &=& - \left(L_{\alpha}/\phi_{\alpha}\right){\bm \nabla}{\mu}_{\alpha}.\quad
  \label{eq:v_dif}
\end{eqnarray}
Neither the mean velocity ${\bm v}$ nor the stress tensor ${\bm \sigma}$ is calculated here.\newline
(II) {\it Viscous} case. In addition to case (I), the hydrodynamic contribution is taken into account;
\begin{eqnarray}
  {\bm v}_{\alpha} &=& {\bm v} - \left(L_{\alpha\alpha}/\phi_{\alpha}\right){\bm \nabla}{\mu}_{\alpha}.\quad \label{eq:v_hyd}
\end{eqnarray}
The Stokes equation, Eq. (\ref{eq:stokes}) is used to solve ${\bm v}$, but the elastic contribution, ${\bm \nabla}\cdot{\bm \sigma}$, is excluded.\newline
(III) {\it Viscoelastic} case. The polymer velocity is expressed by Eq. (\ref{eq:vp2}), including all the diffusion, hydrodynamic, and elastic contributions. The solvent velocity is represented by Eq. (\ref{eq:vs2}) which is identical to Eq. (\ref{eq:v_hyd}).
%
% morph3 /Users/admin/Documents/MATLAB/jcp_figs -> fig5
\begin{figure*}
    \includegraphics[scale=0.425]{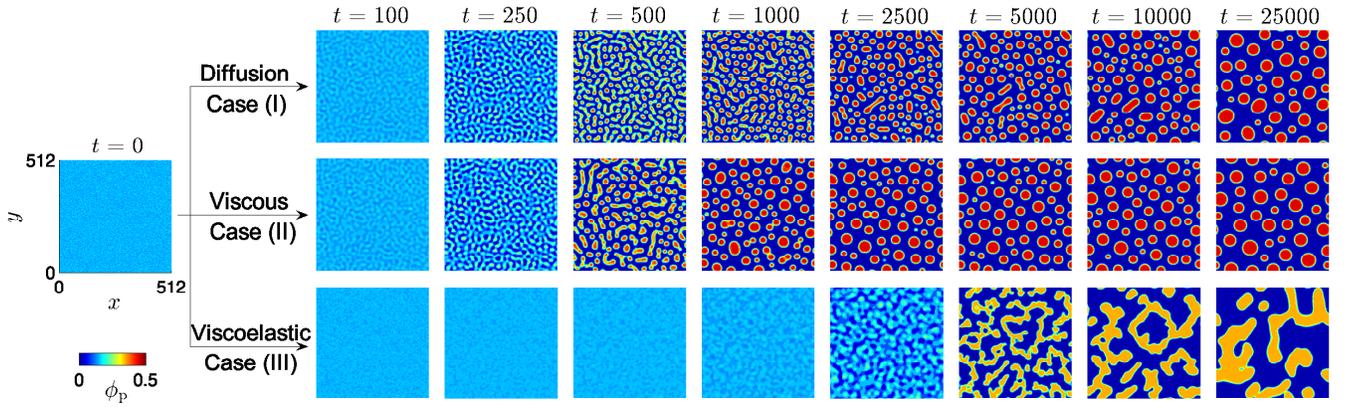}% Here is how to import EPS art
\caption{\label{fig:random3} (color online) Phase separation in the bulk ternary system. Each image represents a two-dimensional distribution of polymer volume fraction $\phi_{\rm p}$ within the simulation box of size $L_{\rm box}\times L_{\rm box}=512^2$. The color is assigned based on the value of $\phi_{\rm p}$, as shown in the scale bar.}
\end{figure*}

\section{Results and Discussion}
\subsection{\label{sec:sd} Phase separation in ternary mixtures}
Figure \ref{fig:random3} illustrates the distributions of the polymer volume fraction $\phi_{\rm p}$ obtained at different times. All simulations are started from a homogeneous mixture, as shown on the leftmost image. Uniform random noise ranging between $-$0.025 and 0.025 is added to the initial volume fraction, whereas it is too small to be visible on the leftmost image.

In case (I), a bicontinuous structure appears vaguely at $t=100$ and becomes clearer at $t=250$. This stage is referred to as the linear (or early) stage of spinodal decomposition, where the polymer-rich and water-rich phases become denser while maintaining their spatial frequencies. Once the polymer-rich phase is broken into pieces through further densification and local shrinkage ($t=500$), smaller pieces of the polymer-rich phase are gradually absorbed into larger domains by an evaporation-condensation mechanism ($t>500$).

In case (II), the morphology change observed in the early stage of spinodal decomposition is almost identical to that in case (I). A noticeable difference can be found at $t=500$, where the polymer-rich domains become larger and more rounded than those in case (I). This is because the deformation of the polymer-rich domains is accelerated by hydrodynamic flow (see Sec. \ref{sec:sq} for the details of the hydrodynamic effects). Once the discretization and circularization of the polymer-rich domains is settled ($t=1000$), the hydrodynamic flow is diminished, and therefore, the morphology change is significantly slowed down. Later, the polymer-rich domains are gradually coarsened through the evaporation-condensation mechanism, similar to case (I).

%\textcolor{red}{
In case (III), a bicontinuous structure is clearly seen at $t=2500$,
which is slower than the other two cases by a factor of $10$. It is also
noteworthy that the time scale of this delay is much longer than the
relaxation time $\tau_{\rm 0}$, {\it i.e.}, 10. In the viscoelastic
phase separation, whenever the polymer-rich domain is deformed by the
thermodynamic force, the elastic force is instantaneously generated as a
counterforce. Then the residual and new thermodynamic force acts again
on the polymer-rich domain, generating the new elastic force in the
polymer-rich domain (see Sec. III-B for the detailed mechanisms). Since
this cycle occurs iteratively and continuously throughout the entire
phase separation process, it takes a considerably long time for the
polymer-rich domain to be fully relaxed.
%}
 Meanwhile, the solvent and water are gradually squeezed out from the polymer-rich phase. As a result, a network-like structure is formed with irregularly shaped polymer-rich domains ($t=5000$). Some frameworks of the network-like structure are maintained over time, whereas others are merged into larger ones ($t>5000$). The latter is driven by a hydrodynamic flow that is caused by disconnection of sharper edges and/or thinner parts of the network-like structure.
%
% morph3 /Users/admin/Documents/MATLAB/jcp_figs -> fig11
\begin{figure}
    \includegraphics[scale=0.3]{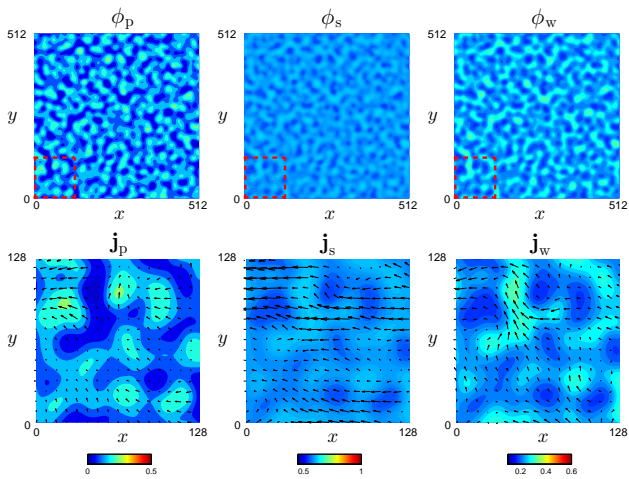}%
\caption{\label{fig:flux3} (color online) Morphology and flow of each component at an early stage of phase separation in case (III) ($t=2500$): (top) two-dimensional distribution of the volume fraction $\phi_{\alpha}$ ($\alpha=$ p,s,w), and (bottom) volumetric flux ${\bm j}_{\alpha}$ within the area enclosed by dashed lines in the top image. A color is assigned to each grid point based on the value of $\phi_{\alpha}$. Arrows on the bottom images represent a vector field of ${\bm j}_{\alpha}$, whose lengths are magnified by $3\times 10^3$ for clarity.}
\end{figure}
%
%\textcolor{red}{
In all three cases, the water-rich phase is formed simultaneously with the polymer-rich phase. As illustrated in Fig. \ref{fig:flux3}, the water distributes inversely to the polymer, reflecting the fact that the water is strongly repelled by the polymer ($\chi_{\rm wp}=2.3$). The water flows in the direction of thickening the water-rich phase, similar to the polymer moving towards the thicker domains.
On the other hand, the solvent distributes over the system and moves across the polymer-rich and water-rich phases. This is mainly because the affinity between the solvent and water ($\chi_{\rm sw}=0.3$) is almost the same as that between the solvent and polymer ($\chi_{\rm ps}=0.4$).
A similar flow trend can be observed at $t = 250$ in cases (I) and (II)
(results are not shown here due to space restriction). Although the
solvent's free movement does not change any basic features of the phase
separation, it becomes particularly important for the NIPS process where
the solvent needs to move across the interface between the polymer
solution and the water\cite{matsuyama99,tree17,tree18,zhou06,garcia20}.
%}

In the rest of the paper, we focus on the analysis of polymer-rich phases to clarify some important roles of polymer viscoelasticity in phase separation in ternary systems.

\begin{figure}
    \includegraphics[scale=0.35]{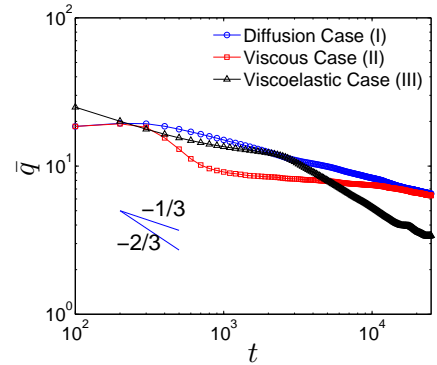}%
\caption{\label{fig:qt} (color online) Change in the characteristic
 wavenumber $\overline{q}$ over time $t$.}
\end{figure}

The morphology change can be quantified with the characteristic
wavenumber $\overline{q}$ defined as\cite{tanaka06}
\begin{eqnarray}
  \overline{q}
   &=& \frac{\int d{\bm q} \left|{\bm q}\right| S\left({\bm q}\right)}{\int d{\bm q}S\left({\bm q}\right)},\label{eq:q}
\end{eqnarray}
where ${\bm q}$ is the wave vector. The structure factor
$S\left({\bm q}\right)$ is obtained
from the square of the Fourier transform of the polymer's volume
fraction, i.e., $\langle |\hat{\phi}_{\rm p}({\bm q})|^2 \rangle$,
where $\langle (\cdots)\rangle$ stands for the statistical average of $(\cdots)$.
Figure \ref{fig:qt} shows $\overline{q}$ of the polymer morphology
sampled over a time period of
$2.5 \times 10^4$ (including the images in Fig. \ref{fig:random3}).
In case (I), $\overline{q}$ is almost constant at $t<300$, and then it
decreases monotonically as
$\overline{q} \propto t^{-1/3}$. The former corresponds to the initial
stage of spinodal decomposition,
and the latter indicates the domain growth driven by the evaporation-condensation mechanism.\cite{taniguchi96}
In case (II), a sharp decrease in $\overline{q}$ is observed from
$t=300$ to $t=800$,
with a rate of $\overline{q} \propto t^{-2/3}$.
%\textcolor{red}{
The exponent of $-2/3$ has also been observed in the binary polymer
solution\cite{taniguchi96} and fluid mixture\cite{furukawa97} where the
numerical simulations were performed in two dimensions including the
hydrodynamic effects. Note, however, that the hydrodynamic-driven domain
growth generally scales with $t^{-1}$.\cite{siggia79,koga93} This
discrepancy is possibly due to the difference in dimension, the
limitation of system size and time scale, or the combination of the
hydrodynamic and diffusion effects\cite{siggia79,scholten05}.
%}
In the later time, $\overline{q}$ remains almost unchanged with the
disappearance of the hydrodynamic flow.
At $t>10^4$, $\overline{q}$ follows the same trend as case (I),
indicating that the polymer-rich domains regrow under the evaporation-condensation mechanism.
In case (III), $\overline{q}$ shows a slight decrease over time,
followed by a sharp drop as
$\overline{q} \propto t^{-2/3}$ at $t > 3000$.
The exponent of $-2/3$ indicates the domain growth driven by the
hydrodynamic flow.
The hydrodynamic flow in case (III) is induced
after breaking off some unstable parts in the network-like structure.
%
% qt3 /Users/admin/Documents/MATLAB/jcp_figs -> fig6
%
%
% kmax3 /Users/admin/Documents/MATLAB/jcp_figs -> fig7
\begin{figure}
    \includegraphics[scale=0.4]{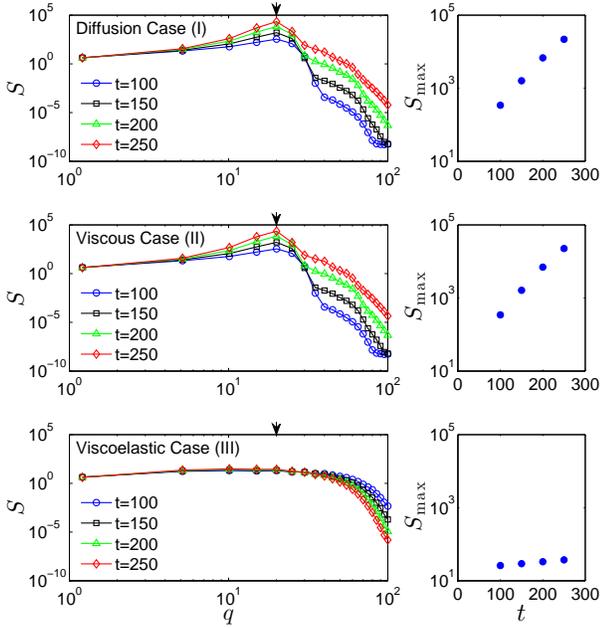}%
\caption{\label{fig:kmax} (color online) Structure factor $S$ at the early stage of spinodal decomposition. (Left) A log-log plot of $S$ as a function of the wavenumber $q$. The arrow points $q=20$ (=$q_{\rm max}$), where $S$ reaches the maximum, $S_{\rm max}$, in cases (I) and (II). (Right) A semilog graph of $S_{\rm max}$ as a function of time $t$.}
\end{figure}

Figure \ref{fig:kmax} illustrates a time evolution of the circularly averaged structure factor $S(q)$ ($q$: wavenumber) at an early stage of spinodal decomposition, calculated from
\begin{eqnarray}
  S\left(q\right) &=& \frac{1}{2 \pi q dq}
    \int_{|{\bm q}|=q}^{|{\bm q}|=q+dq} d{\bm q} ~
    \langle \bigr| \hat{\phi}_{\rm p}\left({\bm q}\right) \bigr |^2 \rangle.
\end{eqnarray}
In cases (I) and (II), $S(q)$ shows a peak at $q_{\rm max} \approx 20$. The peak height $S_{\rm max}$ increases exponentially over time, which is one of the typical characteristics in the early stage of spinodal decomposition. Note that $q_{\rm max}$ can be estimated numerically by solving the eigenequations derived in Appendix \ref{app:kmax}. In case (III), no sharp peak is formed, indicating that the phase separation is suppressed by the elastic effects. Another feature of case (III) is that $S(q)$ decreases over time at $q > q_{\rm max}$. This is due to smoothing of the random noise added to the initial volume fractions. An opposite trend is seen in cases (I) and (II) since the random noise is already diminished and the interface is sharpened over time.

\begin{figure}
    \includegraphics[scale=0.35]{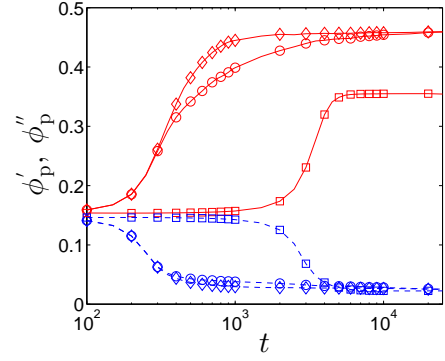}%
\caption{\label{fig:phip3} (color online) Averaged polymer volume fractions in the polymer-rich phase (${\phi}_{\rm p}^{'}$) and in the water-rich phase ($\phi_{\rm p}^{''}$): circle for diffusion case (I), diamond for viscous case (II), and square for viscoelastic case (III). The solid and dashed lines correspond to $\phi_{\rm p}^{'}$ and $\phi_{\rm p}^{''}$, respectively.}
\end{figure}

In addition to the morphological change, the phase compositions also change differently in the three cases. Figure \ref{fig:phip3} illustrates the polymer volume fraction averaged over the polymer-rich domains ($\phi_{\rm p} > \phi_{\rm p0}$) and the polymer volume fraction averaged over the water-rich domains ($\phi_{\rm p}<\phi_{\rm p0}$), denoted by $\phi'_{\rm p}$ and $\phi''_{\rm p}$, respectively. The interfacial regions are excluded from the averaging. In the early stage of spinodal decomposition, $\phi'_{\rm p}$ and $\phi''_{\rm p}$ of case (I) overlap with those of case (II). In the later stage, $\phi'_{\rm p}$ and $\phi''_{\rm p}$ in case (I) asymptotically approach the equilibrium values, $\phi'_{\rm p,eq}$ and $\phi''_{\rm p,eq}$, respectively (see Fig. \ref{fig:equi_comp} for the values of $\phi'_{\rm p,eq}$ and $\phi''_{\rm p,eq}$). On the other hand, in case (II), the phase compositions reach equilibrium almost ten times faster than in case (I). In case (III), $\phi'_{\rm p}$ and $\phi''_{\rm p}$ are constrained to the initial value $\phi_{\rm p0}$ (=0.15) over a relatively long period, and then they exhibit a sharp change similar to those in the early stage of cases (I) and (II).
All these trends are consistent with those observed in Figs. \ref{fig:random3}-\ref{fig:kmax}, except that in case (III), $\phi'_{\rm p}$ settles into a lower value (0.35) than $\phi'_{\rm p,eq}$ (0.48). This can also be seen as a color difference in the polymer-rich domains at $t > 2500$ in Fig. \ref{fig:random3}. The less densified polymer-rich domains are induced by the elastic force, which always acts against the thermodynamic force (see next section for the details of the two forces).
%\textcolor{red}{
It should also be noted that in the plateau regime of case (III),
$\phi'_{\rm p}$ still keeps increasing at a considerably slow rate; it
will take an extremely long time for $\phi'_{\rm p}$ to reach to
$\phi'_{\rm p,eq}$.
%}
%
% morph3 /Users/admin/Documents/MATLAB/jcp_figs -> fig11

\subsection{\label{sec:sq} Driving forces for morphology change}
%
% sq3 /Users/admin/Documents/MATLAB/jcp_figs -> fig8
% manually changed position of colorbar
\begin{figure}
    \includegraphics[scale=0.39]{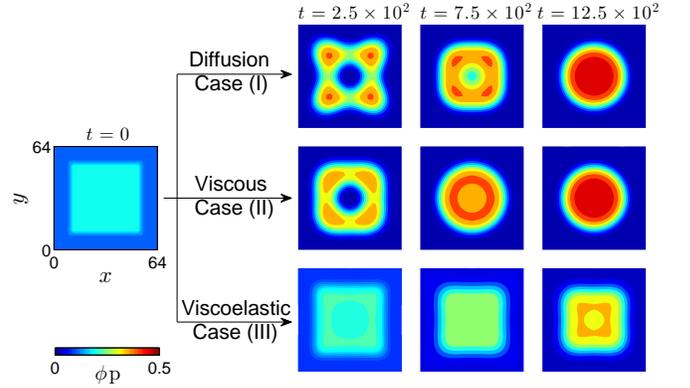}%
\caption{\label{fig:sq3} (color online) Change in shape of the polymer-rich phase. For all cases, the polymer-rich phase is set to be a square at $t = 0$, as shown on the leftmost image. Each image represents a two-dimensional distribution of $\phi_{\rm p}$ within the simulation box of $L_{\rm box} \times L_{\rm box} = 64^2$. Colors are assigned based on the scale bar.}
\end{figure}
%
% v3 /Users/admin/Documents/MATLAB/jcp_figs -> fig9
\begin{figure}
    \includegraphics[scale=0.4]{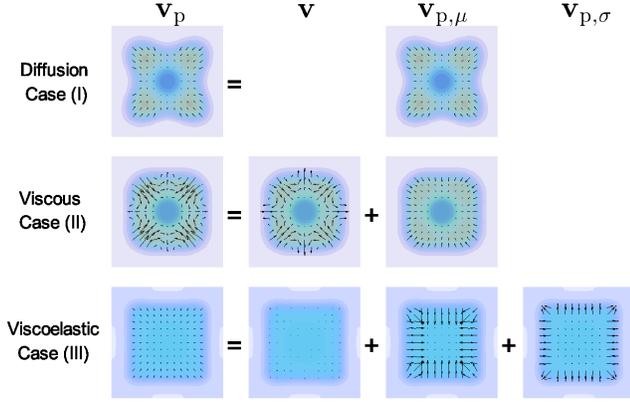}%
\caption{\label{fig:v3} (color online) Polymer velocity ${\bf v}_{\rm p}$ in the polymer-rich phase at $t=2500$ in Fig. \ref{fig:sq3}. The velocity is represented by arrows on top of the distribution of $\phi_{\rm p}$. The velocity field on the leftmost image corresponds to ${\bf v}_{\rm p}$, and the ones on the second, third, and fourth left columns represent the mean velocity ${\bf v}$ (not calculated in case (I)), the polymer velocity driven by the gradient of chemical potential, ${\bf v}_{{\rm p},{\mu}}$, and that driven by the divergence of the stress tensor, ${\bf v}_{{\rm p},{\sigma}}$ (calculated only in case (III)), respectively. For clarity, the number of arrows is reduced from $32\times32$ to $16\times16$, and only those located inside of the polymer-rich phase ($\phi_{\rm p} > 0.15$) are shown here. All arrows are magnified uniformly by 200; however, some of them are still too small to be apparent.}
\end{figure}
To clarify a relationship between the morphology change of the polymer-rich domains and the polymer velocity ${\bm v}_{\rm p}$, we set up another initial condition, as illustrated in the leftmost image in Fig. \ref{fig:sq3}. The compositions of the polymer-rich and water-rich phases, $\phi'_{\alpha}$ and $\phi''_{\alpha}$ ($\alpha =$ p, s, w), are estimated from the bicontinuous structure formed at $t=2500$ in case (III): ($\phi'_{\rm p}$, $\phi'_{\rm s}$, $\phi'_{\rm w}$) = (0.194, 0.584, 0.222), and ($\phi''_{\rm p}$, $\phi''_{\rm s}$, $\phi''_{\rm w}$) = (0.102, 0.617, 0.281).
The size of the square-shaped polymer-rich phase is determined in a way that the averaged volume fractions over the system become equal to the initial volume fractions used in the previous section, $\phi_{\rm \alpha 0}$. With this setting, $\phi'_{\rm p}$ and $\phi''_{\rm p}$ eventually reach $\phi'_{\rm p,eq}$ and $\phi''_{\rm p,eq}$, as seen in Fig. \ref{fig:phip3}.

\begin{figure}
    \includegraphics[scale=0.4]{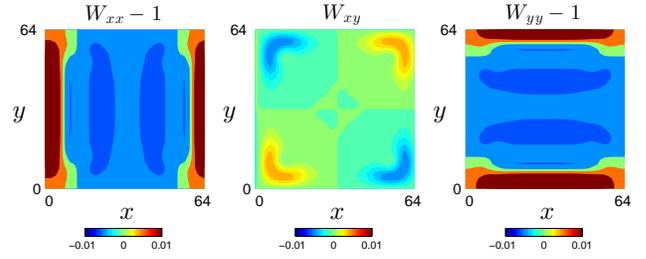}%
\caption{\label{fig:w3} (color online) Conformation tensor ${\bm W}$ at $t=2500$: (left) $W_{xx}-1$, (center) $W_{xy}$, and (right) $W_{yy}-1$. A color is assigned using the scale bar on the bottom side. The negative (or positive) regions in $W_{ii}-1$ ($i=x$ or $y$) indicate where the polymer is locally compressed (or stretched) along the $i$ direction.}
\end{figure}

Figure \ref{fig:sq3} shows the polymer morphology at $t=0$, 250, 750, and 1250 for all three cases.
In case (I), the polymer initially moves towards the corners of the square since the solvent and water diffuse out more from the corners than from the middle edges. As the corners of the square become rich in polymer, the interfaces become concave, and the inside becomes hollow ($t=250$). The polymer movement is plotted in Fig. \ref{fig:v3}, where the majority of the polymer velocity ${\bf v}_{\rm p}$ points to the polymer-rich corner regions. Note that in case (I), ${\bf v}_{\rm p}$ is equal to the velocity driven by the gradient of the polymer potential, ${\bf v}_{\rm p,\mu} (= -L_{\rm pp}/\phi_{\rm p}{\bm \nabla}\mu_{\rm p})$, as defined in Eq. (\ref{eq:v_dif}). The polymer-rich phase constantly exudes the solvent and water until it reaches equilibrium. As a result, the polymer-rich phase is continuously shrunk and transformed into a circular shape ($t=1250$) through a rounded square shape ($t=750$).
In case (II), the morphology change occurs similarly to that in case (I); however, the corners of the polymer-rich square become rounded earlier than in case (I). This is mainly due to the hydrodynamic flow, which is not considered in case (I). As defined in Eq. (\ref{eq:v_hyd}), the polymer velocity ${\bf v}_{\rm p}$ in case (II) is composed of the mean velocity ${\bf v}$ and the diffusion-driven velocity ${\bf v}_{\rm p, \mu}$. For instance, each velocity at $t=250$ can be seen in Fig. \ref{fig:v3}. Similar to case (I), ${\bf v}_{\rm p,\mu}$ points to the polymer-rich corners. On the other hand, ${\bf v}$ exhibits a large flow that enhances the rounding of the polymer-rich phase. Once the polymer-rich phase is transformed into a circular shape ($t=750$), the hydrodynamic effect is diminished, and the change in the composition of the polymer-rich phase is governed by the same diffusion process as in case (I).
In case (III), the square shape of the polymer-rich phase is maintained even at $t=12500$. As expressed in Eq. (\ref{eq:vp2}), ${\bf v}_{\rm p}$ in case (III) is the sum of ${\bf v}$, ${\bf v}_{\rm p, \mu}$, and the velocity driven by the divergence of the stress tensor, ${\bf v}_{\rm p,\sigma} (= L_{\rm pp}/\phi_{\rm p}^2{\bm \nabla}\cdot{\bm \sigma})$. Notably, the flow of ${\bf v}_{\rm p,\sigma}$ in Fig. \ref{fig:v3} is almost perfectly opposite to that of ${\bf v}_{\rm p,\mu}$. This can be interpreted as a balance between the externally acting thermodynamic force and the internally generated elastic force. Indeed, the polymer-rich phase is under internal compression, as shown in Fig. \ref{fig:w3}; the elastic force is generated to counteract
%Editor: Please ensure that the intended meaning has been maintained in this edit.
the compressive strain. Note that according to the Stokes equation as shown in Eq. (\ref{eq:stokes}), ${\bf v}$ is induced only by solvent movement when ${\bf v}_{\rm p, \mu}$ is completely offset
%Editor: Please ensure that the intended meaning has been maintained in this edit.
by ${\bf v}_{\rm p, \sigma}$. Therefore, ${\bf v}$ in case (III) becomes much smaller than in case (II), resulting in a large reduction in any polymer movements.
%
% w3 /Users/admin/Documents/MATLAB/jcp_figs -> fig10

\begin{figure}
    \includegraphics[scale=0.36]{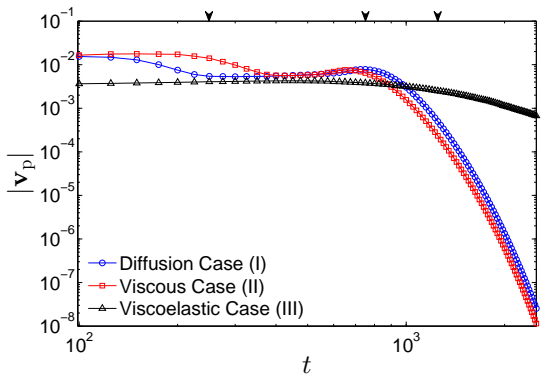}% Here is how to import EPS art
\caption{\label{fig:vp3} (color online) Change in the magnitude of the polymer velocity, $|{\bf v}_{\rm p}|$, over time $t$. The arrows on the top $x$-axis indicate where the data of Fig. \ref{fig:v3} are sampled.}
\end{figure}

The magnitude of the polymer velocity, $\left|{\bf v}_{\rm p}\right|$, as shown in Fig. \ref{fig:vp3}, is calculated by taking an average of the magnitudes of the polymer velocity within the polymer-rich phase ($\phi_{\rm p} > 0.15$). In case (I), $\left|{\bf v}_{\rm p}\right|$ in the early stage shows a gradual decrease over time ($100\le t \le 250$), subsequently remaining almost unchanged ($250 \le t \le 750$). The former corresponds to a slowdown of the polymer movement towards the corners, and the latter corresponds to a shrinkage of the polymer-rich phase driven by diffusion out of the solvent and water. The composition of the polymer-rich phase reaches equilibrium during a sharp drop of $\left|{\bf v}_{\rm p}\right|$ ($ t \ge 750$). In case (II), two plateaus are observed: one from $t=100$ to $t=200$ and another from $t=400$ to $t=600$. The first plateau is associated with the rounding of the polymer-rich phase driven by hydrodynamic flow. The second plateau overlaps with $\left|{\bf v}_{\rm p}\right|$ in case (I), indicating that the morphology change of the polymer-rich phase is driven by the diffusion process, similar to case (I). In case (III), the initial value of $\left|{\bf v}_{\rm p}\right|$ is approximately 5 times smaller than that of cases (I) and (II) due to the elastic effects mentioned above. Since the elastic force is constantly generated to balance out the thermodynamic force, $\left|{\bf v}_{\rm p}\right|$ remains the same order of magnitude over the entire simulation time. This explains why the morphology of the polymer-rich phase remains square, keeping the polymer volume fraction relatively low inside.
%
% v3 /Users/admin/Documents/MATLAB/jcp_figs -> fig11

\section{Conclusion}
We constructed a new model for the viscoelastic phase separation of ternary polymer solutions consisting of entangled polymers, solvents, and nonsolvents. The effects of both elasticity and relaxation of the entangled polymer were incorporated into the model by adding elastic energy to the free energy of the bulk ternary system and by introducing relaxation time into the constitutive equation. Our numerical simulations of the bulk ternary polymer solution demonstrated that the polymer-rich phase was frozen at the early stage of viscoelastic phase separation, whereas it was transformed into a network-like structure and gradually relaxed over a long time at the late stage. Both phenomena
%Editor: Please ensure that the intended meaning has been maintained in this edit.
were essentially the same as those observed in the binary polymer solution. A new finding was that the solvent, which was soluble to the polymer and the nonsolvent, moved across the polymer-rich and water-rich phases during the viscoelastic phase separation of the polymer ternary solution.
%\textcolor{red}{
As a final note, we simplified our model by setting some model
parameters ({\it e.g.}, mobility coefficients $L_{\alpha\beta}$ and the
viscosity $\eta$) to be constant to clarify the viscoelastic effects of
the entangled polymer on ternary phase separation. In reality, these
parameters vary with the concentration and affect the phase separation
to some degree. Indeed, some previous studies showed in simulations that
the concentration dependency of $L_{\alpha\beta}$ and $\eta$ could be
crucial for the formation of the glassy polymer-rich
domains\cite{barton98} and asymmetric polymeric membranes formed in the
NIPS process\cite{tree17,tree18,garcia20}. It should be emphasized here
that both viscous and elastic features of the polymer solution are
fundamentally essential for the viscoelastic phase separation; a simple
addition of the concentration-dependent parameters into the diffusion or
viscous model may not be sufficient to reproduce the frozen and
recovering features of the viscoelastic phase separation
simultaneously. We are currently investigating the combinatory effect of
the viscoelasticity of the entangled polymer, the
concentration-dependent model parameters, and the thermal fluctuations
on the phase separation of ternary polymer solutions.
%}
\begin{acknowledgments}
This work was partially supported by JSPS KAKENHI Grant Number 19H01862. We also acknowledge the supercomputer systems of the Institute for Chemical Research and Academic Center for Computing and Media Studies, Kyoto University.
\end{acknowledgments}

\bigskip
\noindent
{\bf Data Availability}\\
The data that supports the findings of this study are available within the article.

\nocite{*}
%\bibliography{jcp_bib2}% Produces the bibliography via BibTeX.
\bibliography{JCP_Yoshimoto_Taniguchi_revised.bib}% Produces the bibliography via BibTeX.

%%% Appendix %%%
\appendix

\section{\label{app:mu}Chemical potential}
The chemical potentials $\mu_{\rm p}$ and $\mu_{\rm s}$ are expressed as
\begin{eqnarray}
  \mu_{\rm p} = \mu_{\rm p}^* - \mu_{\rm w}^*, \quad \mu_{\rm s} = \mu_{\rm s}^* - \mu_{\rm w}^*, \label{app:eq:mu}
\end{eqnarray}
where
\begin{eqnarray}
  \mu_{\rm p}^*&=&\frac{1}{N}\left(\mbox{ln}\phi_{\rm p}+1\right)+\chi_{\rm ps}\phi_{\rm s}+\chi_{\rm wp}\phi_{\rm w} \nonumber\\
  && - \kappa_{\rm ps}\left(\Delta\phi_{\rm p}-\Delta\phi_{\rm s}\right) - \kappa_{\rm wp}\left(\Delta\phi_{\rm p}-\Delta\phi_{\rm w}\right) \nonumber\\
  && + \frac{3}{4}G_0{\phi_{\rm p}}^2\left({\bm W} - {\bm I}\right):\left({\bm W} - {\bm I}\right), \label{app:eq:mup}\\
  \mu_{\rm s}^*&=&\mbox{ln}\phi_{\rm s} + 1 + \chi_{\rm ps}\phi_{\rm p}+\chi_{\rm sw}\phi_{\rm w} \nonumber\\
  && - \kappa_{\rm ps}\left(\Delta\phi_{\rm s}-\Delta\phi_{\rm p}\right) - \kappa_{\rm sw}\left(\Delta\phi_{\rm s}-\Delta\phi_{\rm w}\right), \label{app:eq:mus}\\
  \mu_{\rm w}^*&=&\mbox{ln}\phi_{\rm w} + 1 + \chi_{\rm sw}\phi_{\rm s}+\chi_{\rm wp}\phi_{\rm p} \nonumber\\
  && - \kappa_{\rm sw}\left(\Delta\phi_{\rm w}-\Delta\phi_{\rm s}\right) - \kappa_{\rm wp}\left(\Delta\phi_{\rm w}-\Delta\phi_{\rm p}\right). \label{app:eq:muw}
\end{eqnarray}

\section{\label{app:kmax} Analysis of the characteristic wavelength in ternary spinodal decomposition}
\begin{figure}
    \includegraphics[scale=0.5]{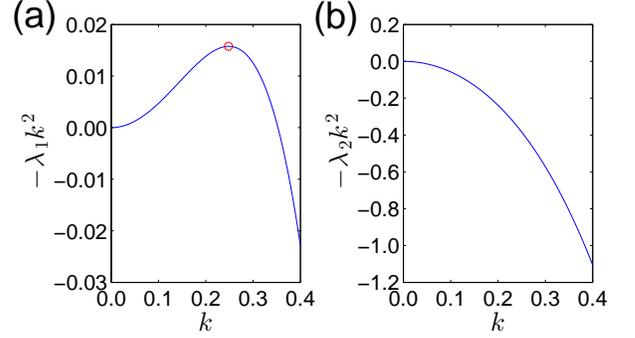}% Here is how to import EPS art
\caption{\label{fig:lambda3} Frequency-dependent coefficient $-\lambda k^2$: (a) $-\lambda_1k^2$ and (b) $-\lambda_2k^2$. The circular mark on (a) denotes the maximum point, (0.248, 0.0158).}
\end{figure}

Here, we assume that ternary phase separation occurs by a simple diffusion mechanism, as in case (I). The continuity equation for the polymer is written as
\begin{eqnarray}
  \frac{\partial \phi_{\rm p}}{\partial t} &=& L_{\rm pp}{\bm \nabla}\cdot\left[{\bm \nabla}\frac{\delta F}{\delta \phi_{\rm p}}\right]\nonumber\\
  &=& L_{\rm pp}\Delta\left(h_{\rm p} + K_{\rm p1}\Delta\phi_{\rm p} + K_{\rm p2}\Delta\phi_{\rm s}\right),\label{eq:chp}
\end{eqnarray}
where
\begin{eqnarray}
  h_{\rm p} &\approx& \left.\frac{\partial f_{\rm mix}}{\partial \phi_{\rm p}}\right|_{\phi_{\alpha 0}} + \left.\frac{\partial^2 f_{\rm mix}}{\partial \phi_{\rm p}^2}\right|_{\phi_{\alpha 0}}\left(\phi_{\rm p} - \phi_{\rm p0}\right)\label{eq:hp},\\
   K_{\rm pp} &=& - \kappa_{\rm ps} - \kappa_{\rm sw} + 4\kappa_{\rm wp}\label{eq:kp1},\\
   K_{\rm ps} &=& \kappa_{\rm ps}^2 - \kappa_{\rm sw} + 2\kappa_{\rm wp}\label{eq:kp2}.
\end{eqnarray}
At the early stage, each volume fraction deviates slightly from the initial values;
\begin{eqnarray}
  &&\phi_{\alpha}({\bm r},t) = \phi_{\alpha 0} + u_{\alpha}({\bm r},t),\label{eq:u}
\end{eqnarray}
where $u_{\alpha}({\bm r},t)$ ($\alpha = $p, s) denotes a small perturbation.
By substituting Eq. (\ref{eq:u}) into Eq. (\ref{eq:chp}), the polymer continuity equation is expressed with respect to $u_{\rm p}$;
\begin{eqnarray}
  \frac{\partial u_{\rm p}}{\partial t} = L_{\rm pp}\Delta\left(\left.\frac{\partial^2 f_{\rm mix}}{\partial \phi_{\rm p}^2}\right|_{\phi_{\alpha 0}}u_{\rm p} + \sum_{\alpha={\rm p,s}}K_{\rm p\alpha}{\Delta}u_{\alpha}\right).\label{eq:chp2}
\end{eqnarray}
Similarly, the continuity equation for the solvent can be described as
\begin{eqnarray}
  \frac{\partial u_{\rm s}}{\partial t} = L_{\rm ss}\Delta\left(\left.\frac{\partial^2 f_{\rm mix}}{\partial \phi_{\rm s}^2}\right|_{\phi_{\alpha 0}}u_{\rm s} + \sum_{\alpha={\rm p,s}}K_{\rm s\alpha}{\Delta}u_{\alpha}\right),\label{eq:chs2}
\end{eqnarray}
where
\begin{eqnarray}
   K_{\rm sp} &=& \kappa_{\rm ps}^2 - 2\kappa_{\rm sw}^2 + 2\kappa_{\rm wp}^2,\label{eq:ks1}\\
   K_{\rm ss} &=& -\kappa_{\rm ps}^2 - 4\kappa_{\rm sw}^2 + \kappa_{\rm wp}^2.\label{eq:ks2}
\end{eqnarray}
After taking the Fourier transform, Eqs. (\ref{eq:chp2}) and (\ref{eq:chs2}) can be written in matrix form as
\begin{eqnarray}
  \frac{d}{dt}\left[
    \begin{array}{c}
    \hat{u}_{\rm p}\\
    \hat{u}_{\rm s}
    \end{array}
    \right] = -k^2
  {\bm \Omega}
  \left[
    \begin{array}{c}
    \hat{u}_{\rm p}\\
    \hat{u}_{\rm s}
    \end{array}
    \right]\label{eq:uq}
\end{eqnarray}
where $k$ is the angular frequency (=$2\pi q/L_{\rm box}$) and $\hat{u}_{\alpha}$ is the Fourier transform of $u_{\alpha}$. The matrix ${\bm \Omega}$ is defined as
\begin{eqnarray}
  {\bm \Omega} = 
   \left[ 
    \begin{array}{cc}
      L_{\rm pp}\Bigr ( \left.\frac{\partial^2 f_{\rm mix}}
                         {\partial \phi_{\rm p}^2}\right|_{\phi_{\alpha 0}} 
                       - K_{\rm pp}k^2
                \Bigr )
   &  -L_{\rm pp}K_{\rm ps}k^2
   \\
      L_{\rm ss}\Bigr( \left.\frac{\partial^2 f_{\rm mix}}
                        {\partial \phi_{\rm s}^2}\right|_{\phi_{\alpha 0}}
                       - K_{\rm sp}k^2
                \Bigr)
   & -L_{\rm ss}K_{\rm ss}k^2
    \end{array}
    \right].\quad\quad      
\end{eqnarray}
The eigenvalues of ${\bm \Omega}$, $\lambda$, can be obtained by solving
\begin{eqnarray}
\mbox{det}\left|{\bm \Omega} - {\lambda}{\bm
           I}\right|=0,\label{eq:lambda}
\end{eqnarray}
which can be negative or positive depending on the value of $k$.
In the eigenspace, ${\bm \Omega}$ on the right-hand side of Eq. (\ref{eq:uq}) can be replaced with a diagonal matrix composed of $\lambda$. Therefore, the sign of $-\lambda k^2$ is essential to determine whether fluctuation of the volume fractions at a given frequency $k$ is amplified or diminished over time. The two eigenvalues calculated from Eq. (\ref{eq:lambda}) are plotted  in Fig. \ref{fig:lambda3}. Here, $-\lambda_1k^2$ forms a positive peak at $k=0.248$ ($\equiv k_{\rm max}$), whereas $-\lambda_2k^2$ always becomes negative at $k > 0$. This indicates that the fluctuation of the volume fraction at $k=k_{\rm max}$ may be amplified at the fastest rate. Note that the wavenumber corresponding to $k_{\rm max}$ is 20.2; the result here agrees with $q_{\rm max}$ as shown in Fig. \ref{fig:kmax}.
%
% lambda /Users/admin/Documents/MATLAB/jcp_figs -> fig111

\end{document}